\documentclass{aa}  

\usepackage{graphicx}
\usepackage{txfonts}
\usepackage[colorlinks=true, linkcolor=black, citecolor=blue, urlcolor=black]{hyperref}
\begin{document}

\title{Self-consistent modeling of the energetic storm particle event\\ of 10 November 2012}

\author{A. Afanasiev\inst{1}
        \and
        R. Vainio\inst{1} 
        \and
        D. Trotta\inst{2} 
        \and
        S. Nyberg\inst{1} 
        \and
        N. Talebpour Sheshvan\inst{1}
        \and 
        H. Hietala\inst{3} 
        \and 
        N. Dresing\inst{1}
        }
        
\institute{Department of Physics and Astronomy, University of Turku, Finland\\
            \email{alexandr.afanasiev@utu.fi}
        \and
            The Blackett Laboratory, Department of Physics, Imperial College, London, United Kingdom
        \and
            Department of Physics and Astronomy, Queen Mary University of London, London, United Kingdom
        }

\date{}

\abstract
{It is thought that solar energetic ions associated with coronal/interplanetary shock waves are accelerated to high energies by the diffusive shock acceleration mechanism. In order to be efficient, this mechanism requires intense magnetic turbulence in the vicinity of the shock. The enhanced turbulence upstream of the shock can be produced self-consistently by the accelerated particles themselves via streaming instability. Comparisons of quasi-linear-theory-based particle acceleration models including this process with observations have not been fully successful so far, which has been a reason for the development of acceleration models of different nature.}
{Our aim is to test how well our self-consistent quasi-linear SOLar Particle Acceleration in Coronal Shocks (SOLPACS) simulation code, developed earlier to simulate proton acceleration in coronal shocks, models the particle foreshock region.}
{We apply SOLPACS to model the energetic storm particle (ESP) event observed by the STEREO A spacecraft on November 10, 2012.}
{All but one main input parameters of SOLPACS are fixed by the in-situ plasma measurements from the spacecraft. Comparison of a simulated proton energy spectrum at the shock with the observed one allows us to fix the last simulation input parameter related to efficiency of particle injection to the acceleration process. Subsequent comparison of simulated proton time-intensity profiles in a number of energy channels with the observed ones shows a very good correspondence throughout the upstream region.}
{Our results strongly support the quasi-linear description of the foreshock region.}


\titlerunning{Self-consistent modeling of energetic storm particle event}
\authorrunning{A. Afanasiev et al.}

\maketitle


\section{Introduction} \label{sec:intro}
Shock waves driven by coronal mass ejections (CMEs) are considered to be the source of the so-called gradual solar energetic particle (SEP) events, in which ions can achieve high energies (tens/hundreds of MeV/nuc) \citep[e.g.,][]{Reames17}. Although the general mechanism -- diffusive shock acceleration (DSA) \citep{AxfordLeerSkadron77, Krymskii77, Bell78, BlandfordOstriker78} -- by which ions can be accelerated to such energies is widely accepted, the detailed acceleration process is still under investigation. 

One of the open questions is the origin of the turbulent magnetic fluctuations upstream of the shock, which are required by DSA. Estimations of the DSA efficiency, at least for coronal shocks, suggest that magnetic turbulence should be substantially enhanced close to the shock as compared to its level as derived for the ambient solar wind \citep{NgReames08}. A rather prevalent understanding is that this turbulence is self-generated, i.e., excited by the accelerated particles (mainly protons) themselves via the streaming instability \citep[e.g.,][]{Lee83, ZankRiceWu00, NgREames94, Vainio03}. This mechanism, however, requires the flux of streaming particles to exceed some threshold in order to be efficient \citep{Vainio03}. This causes skepticism regarding the self-generated turbulence resonant with high-energy (hundreds of MeV) protons being able to develop since proton energy spectra in shocks are expected to be steep \citep{KocharovLaitinenVainio13, Kocharov15}.       

The above argument is used to support another possible scenario where intense turbulence is inherent to the solar wind, being present in some magnetic flux tubes and absent in others \citep{KocharovLaitinenVainio13, Kocharov15}. When a shock travels through such a structured solar wind, particles in turbulent flux tubes are efficiently trapped near the shock and accelerated via DSA, whereas particles appearing in quiet tubes are able to escape from the shock vicinity to the ambient solar wind (in this scenario the cross-field transport delivers accelerated particles from turbulent tubes to quiet ones). 

Although a variety of models of ion acceleration in shocks including self-generated (Alfv\'enic) turbulence exist \citep[e.g.,][]{Bell78, Lee83, GordonLeeMobius99, ZankRiceWu00, Lee05, Vainio07, NgReames08, Afanasiev15, BerezhkoTaneev2016, Li22}, it is difficult to evaluate how well they reflect the reality, especially at early times, when the shock is still in the corona. In this respect, it is important to note that they are based on quasi-linear theory of wave-particle interactions\footnote{Here we do not consider self-consistent kinetic models, which are small-scale models in terms of the simulated time and spatial extent of the system.}. However, one can try to acquire a better understanding into this question by studying the so-called energetic storm particle (ESP) events. ESP events are particle intensity enhancements associated with passages of interplanetary (IP) shocks over observing spacecraft \citep[see, e.g., a review by][and references therein]{DesaiGiacalone2016}, so particle intensity measurements in the immediate vicinity of the shock and shock in-situ measurements are available during such events. Therefore ESP events can be used for more accurate tests and validation of existing models of particle acceleration in shocks.

ESP events can be of different types \citep[e.g.,][]{Lario03}. The so-called classic ESP events are characterized by gradual increase of the particle intensity typically starting a few hours prior to the shock arrival to the spacecraft. Such ESP events are qualitatively consistent with the predictions of DSA \citep[e.g.,][]{Giacalone12}.

Attempts to compare DSA modeling including self-generated Alfv\'enic turbulence with some classic type ESP events were made by \citet{Kennel86} and more recently by \citet{BerezhkoTaneev2016} and \citet{Taneev18}. \citet{Kennel86} tested the early analytical model by \citet{Lee83}, which describes particle acceleration using Parker's equation, i.e., it is based on the assumption that the condition of spatial diffusion (strong scattering condition) is satisfied everywhere in the shock upstream. Although \citet{Kennel86} considered the theory predictions for accelerated particles to be rather successful, one can notice (see their Fig.~1) that, e.g., the measured particle intensities tend to fall off with increasing distance from the shock following a power law rather than exponentially as predicted by the model. Noteworthy, \citeauthor{Vainio07}'s (\citeyear{Vainio07}) model predicts a $\sim 1/(1 + x/x_0)$ dependence for the particle intensity vs. distance $x$ from the (parallel) shock towards upstream ($x>0$ and $x_0$ is the energy-dependent length scale) in agreement with \citet{Bell78}. 

The model by \citet{BerezhkoTaneev2016}  represents a modification of the analytical treatment by \citet{GordonLeeMobius99}, which is, in turn, an improvement of \citeauthor{Lee83}'s (\citeyear{Lee83}) model. \citet{BerezhkoTaneev2016} simplify the expression for the Alfv\'en wave growth rate derived by \citet{GordonLeeMobius99}, using an approximate wave-particle resonance condition (omitting the pitch-angle cosine). On the other hand, they include an additional ad hoc factor (of the order of 0.1) in their expression of the growth rate in order to fit the model results to observations. 

The difficulties of the quasi-linear models involving self-generated turbulence in accurately describing ESP events have contributed to the motivation to search for particle acceleration models other than quasi-linear ones. One such model implements anomalous (super-diffusive) particle transport \citep[e.g.,][]{ZimbardoPerri13, Trotta2020, Perri22}. The super-diffusive transport results in a power-law dependence of the particle intensities vs. time in the shock upstream.  

Detailed comparisons of different physical models (quasi-linear, anomalous transport-based) with observed ESP events are needed in order to understand which scenario takes place in a given event. In this work, we present simulations of the ESP event of November 10, 2012 as observed by the STEREO-A spacecraft, using the SOlar Particle Acceleration in Coronal Shocks (SOLPACS) Monte Carlo code \citep{Afanasiev15} upgraded to allow simulations for oblique shocks (i.e., shocks for which the shock normal and the upstream magnetic field are not aligned). SOLPACS is based on the quasi-linear theory of wave-particle interactions. In contrast to the model of \citet{BerezhkoTaneev2016}, SOLPACS simulates particle pitch-angle diffusion instead of spatial diffusion and uses the exact gyro-resonance condition of wave-particle interactions.    

The paper is organized as follows. In section~\ref{sec:code}, some details of the SOLPACS simulation code are provided. Section~\ref{sec:observations} describes the observations of the ESP event and relevant data analysis. Section~\ref{sec:results} presents simulation results and their discussion. Finally, section~\ref{sec:conclusions} contains our conclusions.

\section{Simulation code} \label{sec:code}
SOLPACS is a Monte Carlo simulation code designed to model acceleration of protons in a shock, including the generation of Alfv\'{e}n waves in the upstream region by the accelerated particles themselves. Here, we provide only the key points of the code and details concerning the implementation of oblique shocks, while the other details can be found in \citet{Afanasiev15} \citep[see also][]{Afanasiev13}.   
The code is based on a spatially one-dimensional (1-D) local formulation, i.e., particles and waves are traced under the guiding-center approximation along a single open magnetic field line, and the ambient plasma parameters in the 1-D spatial simulation box along the magnetic field line (the plasma density $n$, the magnetic field magnitude $B$, and the bulk plasma speed $U$) as well as the shock parameters (e.g., the shock speed $V_\mathrm{sh}$, the shock-normal angle $\theta_\mathrm{Bn}$, etc.) are taken to be constant. The spatial simulation box is placed in the upstream of the shock and limited by the MHD shock on one side, while on the other side of the box a free-escape boundary for particles is assumed.   

The equations that SOLPACS solves can be given in the following form:
\begin{align}
    \frac{\partial f}{\partial t}+\left[v\mu+\left(1-M_{\mathrm{A}}\right)V_{\mathrm{A}}\right]\frac{\partial f}{\partial x} &= \frac{\partial}{\partial\mu}\left(D_{\mu\mu}\frac{\partial f}{\partial\mu}\right),\label{particle_eq}\\
    \frac{\partial I}{\partial t}+\left(1-M_{\mathrm{A}}\right)V_{\mathrm{A}}\frac{\partial I}{\partial x} &= \Gamma\,I, 
    \label{wave_eq}
\end{align}
where Eq.~\eqref{particle_eq} describes the evolution of the gyro-averaged distribution function $f(x,v,\mu,t)$ of particles, and Eq.~\eqref{wave_eq} describes the evolution of the Alfv\'en wave intensity $I(x,k,t)$. Here $t$ is time, $x$ is the spatial coordinate measured along the magnetic field line wrt. the shock ($x>0$ towards upstream), $k$ is the wavenumber, $v$ and $\mu$ are the particle speed and the pitch-angle cosine as measured in the rest frame of Alfv\'{e}n waves propagating at speed $V_\mathrm{A}$ wrt. the ambient plasma, and $M_\mathrm{A}$ is the Alfv\'{e}nic Mach number of the shock. The latter quantity is defined here as the ratio of the upstream bulk plasma speed, $u_1$, in the de Hoffmann-Teller (HT) frame\footnote{
The velocity transformation to the HT frame can be done directly from the spacecraft frame \citep{PaschmannDaly98} or from the normal incidence shock-rest frame \citep[e.g.,][]{KivelsonRussell95}. The latter approach is implied in Eq.~\eqref{Alfvenic_Max_number}. In this case the shock speed needs to be determined first, which can be done by assuming conservation of the mass flux across the shock.
}
to the Alfv\'en speed, i.e., 
\begin{equation}
    M_\mathrm{A} = \frac{u_1}{V_\mathrm{A}} = \frac{V_\mathrm{sh}-\hat{\mathbf{n}} \cdot \mathbf{U}}{V_\mathrm{A}\cos{\theta_\mathrm{Bn}}} = \frac{M^{*}_\mathrm{A}}{\cos{\theta_\mathrm{Bn}}},
    \label{Alfvenic_Max_number}
\end{equation}
where $V_\mathrm{sh}$ is the shock speed (measured along the shock normal $\hat{\mathbf{n}}$) and $\mathbf{U}$ is the bulk plasma velocity, both measured in the spacecraft frame, and $M^{*}_\mathrm{A}$ is the conventional Alfv\'enic Mach number.  
Equation~\eqref{particle_eq} represents the spatially 1-D quasi-linear transport equation \citep[e.g.,][]{Skilling75} written in the shock frame, assuming homogeneous magnetic field and plasma conditions; it describes the particle transport upstream of the shock. In this equation, the second term on the left-hand side describes particle streaming wrt. the shock, and the right-hand side term provides the pitch-angle scattering of particles. 
In Eq.~\eqref{wave_eq}, the second term on the left-hand side describes the wave advection towards the shock, and the term on the right-hand side describes the wave growth. The coefficients $D_{\mu\mu}\left(\mu\right)$ and $\Gamma\left(k\right)$ are the quasi-linear pitch-angle diffusion coefficient \citep{Jokipii66} and wave growth rate \citep{Vainio03} respectively: 
\begin{gather}
D_{\mu\mu}\left(\mu\right)=\frac{\pi}{2}\Omega_{0}\frac{\left|k_{\mathrm{r}}\right|I\left(k_{\mathrm{r}}\right)}{\gamma B^{2}}\left(1-\mu^{2}\right)\label{diff_coef} \\    
\Gamma\left(k\right)=\frac{\pi}{2}\frac{\Omega_{0}}{nV_{\mathrm{A}}}\int\mathrm{d^{3}}p\,v\left(1-\mu^{2}\right)\left|k\right|\delta\left(k-k_{\mathrm{r}}\right)\frac{\partial f}{\partial\mu}\label{wave_grw},
\end{gather}
where $\Omega_0$ is the proton cyclotron frequency, $\gamma$ is the relativistic Lorentz-factor, $k_{\mathrm{r}}$ is the resonant wavenumber given by the resonance condition
\[
k_{\mathrm{r}} = \frac{\Omega_0}{\gamma v \mu},
\]
and $I\left(k_{\mathrm{r}}\right)$ is the resonant wave intensity. In Eq.~\eqref{wave_grw}, $\delta\left(\cdot\right)$ is the Dirac delta-function, and the integration is performed over the particle momentum space.  

We assume the plasma turbulence in the box to be due to outward-propagating (if considered in the solar wind frame) Alfv{\'e}n waves, with the initial spectral form $\propto k^{-q_0}$, where $q_0$ is the spectral index. The initial level of turbulence is chosen to provide a prescribed value of the initial mean free path $\lambda_0$ for 100 keV protons. 

The process of seed particle injection into the acceleration process due to their interaction with the shock is not modeled in SOLPACS. As known, in order to address this self-consistently, one has to resort to a kinetic description of the shock \citep[see, e.g.,][and reference therein]{Caprioli2014, Trotta2021}. In SOLPACS, we prescribe a suprathermal proton population being injected at the shock to the upstream throughout the simulation. This population is characterized by the following (shock-frame) velocity spectrum: 
\begin{equation}
\frac{{\rm d}N_{\mathrm{inj}}}{{\rm d}v}=\frac{N_{\mathrm{inj}}}{v_{1}}H(v-u_{1})\mathrm{e}^{-(v-u_{1})/v_{1}},
\label{eq:seed_spectrum}
\end{equation}
where $N_{\mathrm{inj}}=\epsilon_\mathrm{inj}n u_1 t_\mathrm{sim}$ is the total number of suprathermal particles injected into the acceleration process per unit cross-section of the magnetic flux tube during the whole simulation of duration $t_\mathrm{sim}$, $v_{1}$ is a parameter characterizing the extension of the exponential "tail" of the distribution, and $H\left(\cdot\right)$ is the Heaviside step function. The parameter $\epsilon_\mathrm{inj}$, therefore, characterizes the injection efficiency of the shock. It is one of the key SOLPACS parameters controlling the particle acceleration process, since appreciable wave growth at a given point in space requires a finite number of particles to pass through that point \citep{Vainio03, Vainio07}.   

Interaction of injected particles with the MHD shock is treated in the scatter-free approximation and assuming conservation of particle's energy and magnetic moment (in the shock de Hoffmann-Teller frame of reference). Interacted particles can be reflected back to the upstream or transmitted to the downstream region \citep[see, e.g.,][]{Battarbee13}. The reflection/transmission condition is determined by the magnetic compression ratio of the shock, which (like the gas compression ratio) is calculated based on Rankine-Hugoniot conditions. Instead of tracing particles in the shock downstream, we employ a probability of return from the downstream region \citep{JonesEllison91, Vainio07}. 

To run a SOLPACS simulation, input of five main physical parameters is required: the Alfv\'en speed $V_\mathrm{A}$ and density $n$ of the ambient plasma, the Alfv\'enic Mach number of the shock $M_\mathrm{A}$, the shock-normal angle $\theta_\mathrm{Bn}$, and the particle injection efficiency of the shock $\epsilon_\mathrm{inj}$.\footnote{In fact, the upstream plasma beta parameter is needed to be specified in order to calculate the gas compression ratio of the shock based on Rankine-Hugoniot conditions. However, since it often has quite large uncertainty in the observations, we do not consider it as the main input parameter of the code.} 
All these parameters but the last one can be derived from in-situ plasma measurements.

\section{Observations} \label{sec:observations}
\begin{figure*}
\centering
\includegraphics[width=0.8\textwidth]{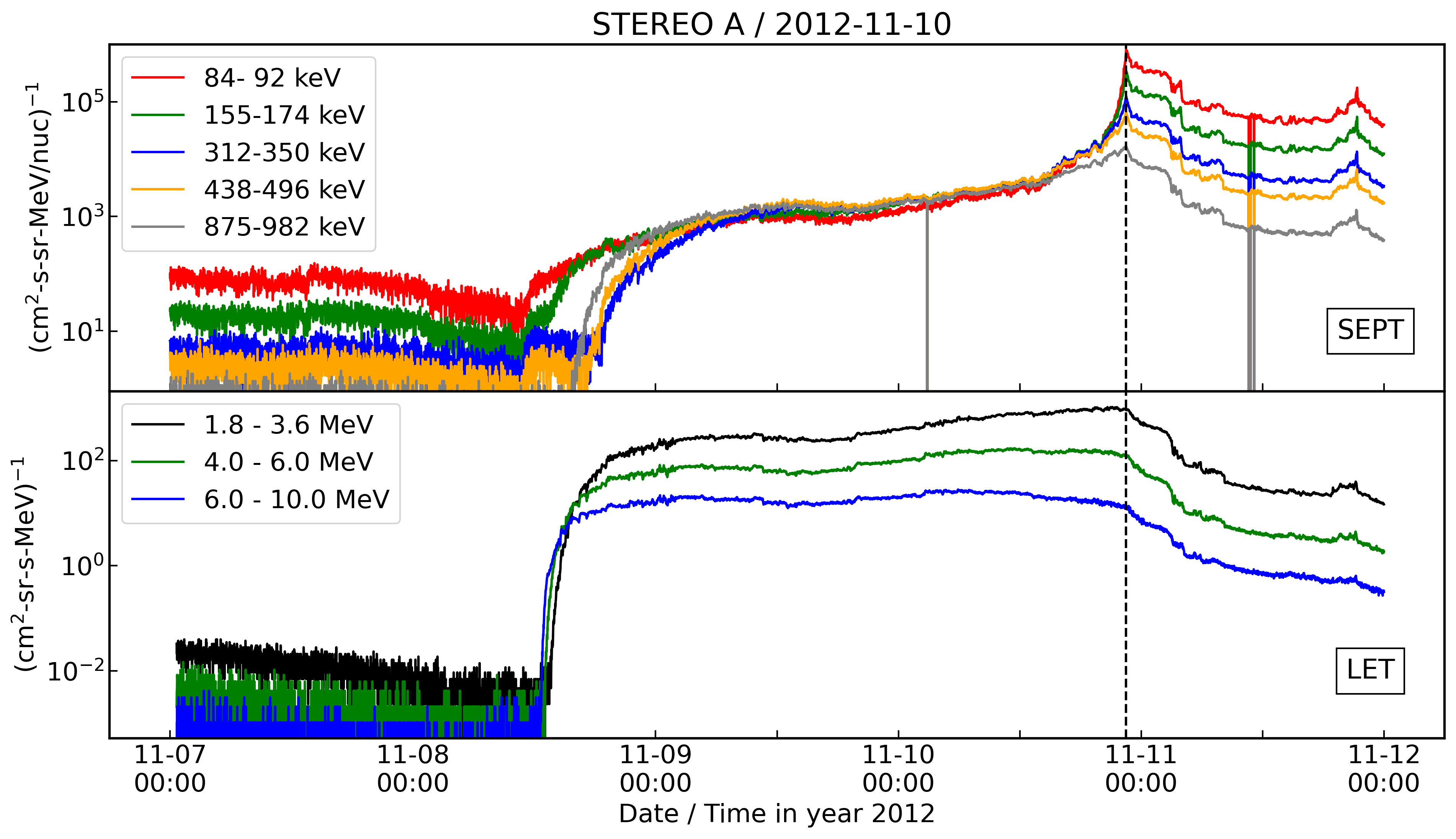}
\caption{Omnidirectional particle intensities vs. time measured in different energy channels by the SEPT (top panel) and LET (bottom panel) instruments onboard the STEREO-A spacecraft, showing the SEP event associated with the modeled ESP event of November 10, 2012. 
The vertical dashed line marks the time of shock arrival at the spacecraft.
\label{fig:sep_data}}
\end{figure*}

\begin{figure*}
\centering
\includegraphics[width=0.8\textwidth]{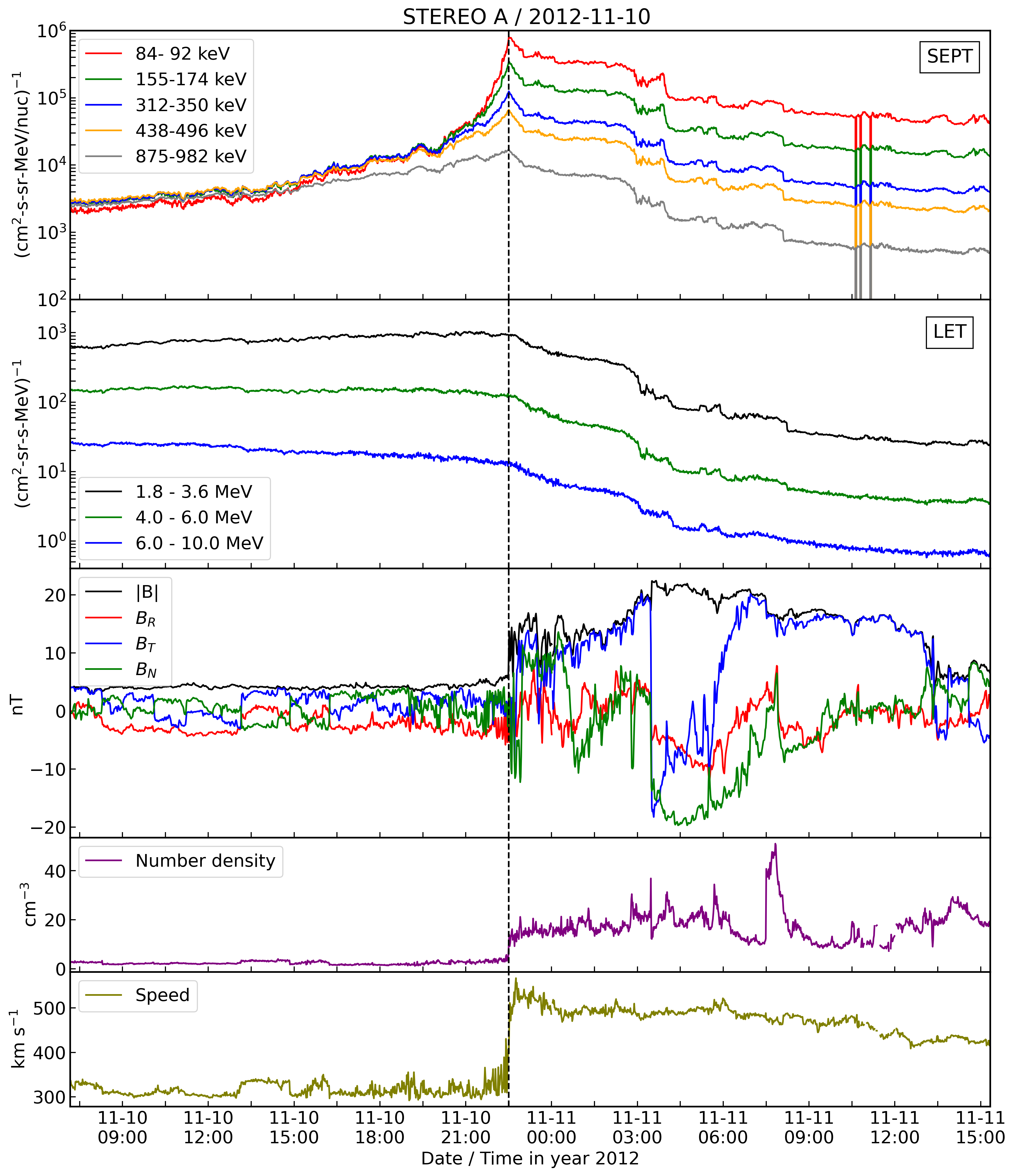}
\caption{Zoom-in around the shock arrival time to STEREO-A, showing (from top to bottom) the particle time-intensity profiles measured by the SEPT and LET instruments, the magnetic field vector measured by the IMPACT/MAG instrument, and the plasma density and the bulk solar wind speed measured by the PLASTIC instrument.  
\label{fig:esp_data_zoom_in}}
\end{figure*}

We used the STEREO IP shock list\footnote{The list can be accessed from 
\url{https://stereo-ssc.nascom.nasa.gov/pub/ins_data/impact/level3/}.} 
to look for clearly visible classic ESP events. From those, we have selected an ESP event occurred on November 10, 2012 for the simulation modeling with SOLPACS. This ESP event is associated with a preceding SEP event (Fig.~\ref{fig:sep_data}) and has clear intensity enhancements peaking at the shock arrival time at energies up to $\sim 1$~MeV (Fig.~\ref{fig:esp_data_zoom_in}). One can see that the pronounced increases in the particle intensities measured by the SEPT instrument start at least four hours before the shock arrival.  

We used the IP shock database\footnote{\url{http://www.ipshocks.fi/}} 
of University of Helsinki to get values for the \textit{in-situ} plasma and shock parameters needed to specify the SOLPACS input parameters. These are presented (together with error estimates) in Table~\ref{table:sim_params}. One can notice a large error obtained for the shock-normal angle $\theta_\mathrm{Bn}$. Besides, the STEREO IP shock list gives for $\theta_\mathrm{Bn}$ a value of $61.5^{\circ}$. The large error and the large difference between the values for $\theta_\mathrm{Bn}$ motivated us to conduct more detailed analysis of the magnetic field time series and calculations of $\theta_\mathrm{Bn}$. 

\begin{figure*}
\centering
\includegraphics[width=0.8\textwidth]{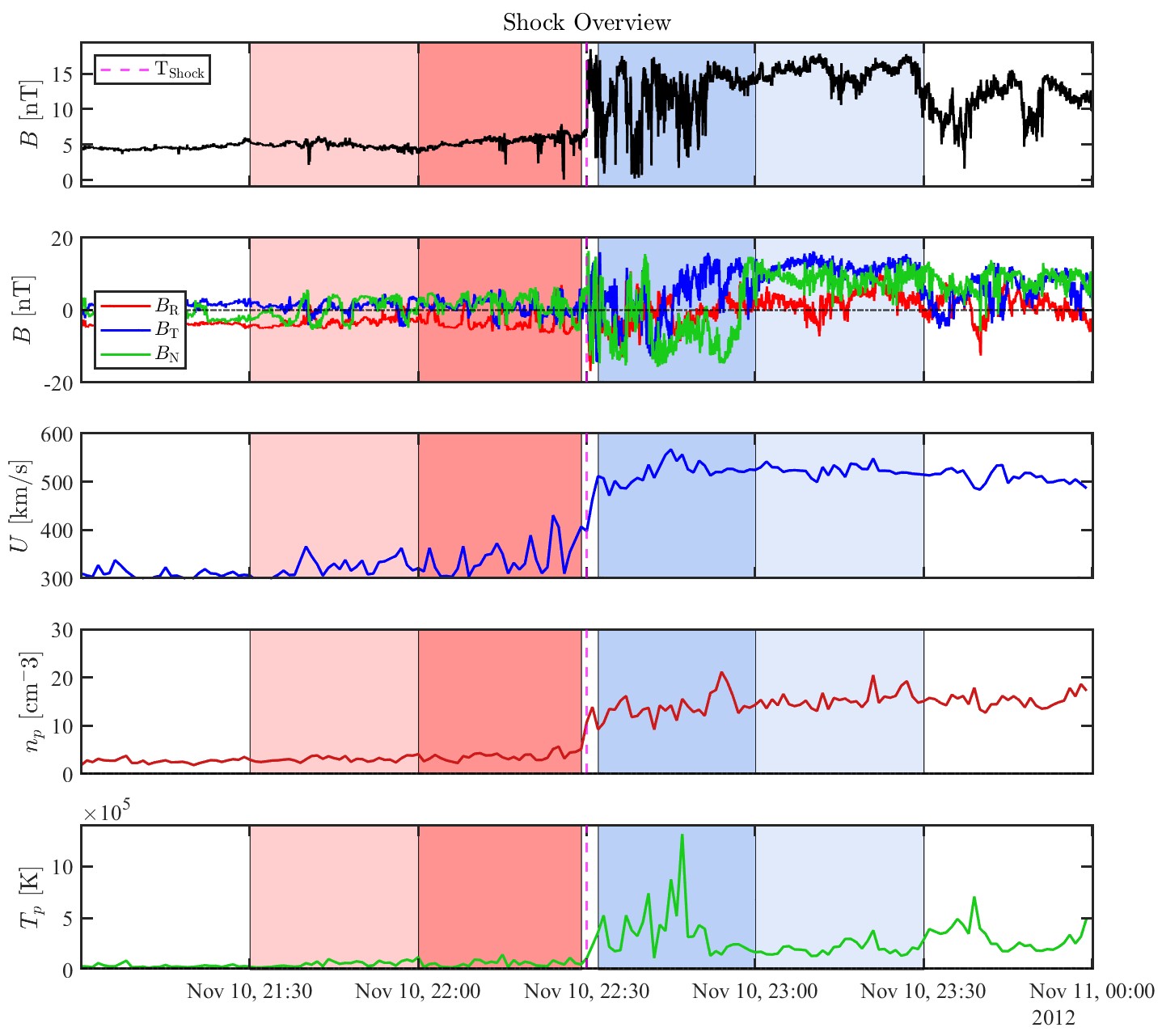}
\caption{Magnetic field magnitude and components, solar wind speed, proton number density, and proton temperature vs. time in the 3-hour time window centered at the shock arrival time. The shock arrival time ($\sim $~22:30~UT) is marked by the dashed magenta line. The red and blue shaded areas denote the shortest (dark) and largest (light) averaging windows used for the shock parameter estimation upstream and downstream, respectively. 
\label{fig:shock_overview}}
\end{figure*}

The magnetic field time series analysis indicates that there are multiple heliospheric current sheet crossings during the 30 min preceding the shock arrival at the spacecraft (Fig.~\ref{fig:shock_overview}). The downstream is even more complex, and the large number of data points with $|\mathbf{B}| \sim 0$~nT indicate that the shock has been propagating through the current sheet for a while. Naturally, the frequent sign flips of $B_\mathrm{R}$ and instances of near zero field magnitude make the shock parameter estimation particularly challenging, due to the high sensitivity to the choice of upstream/downstream averaging time window. This motivated us to look at distribution of $\theta_\mathrm{Bn}$ obtained for a variety of averaging window lengths, i.e., applying the systematic window variation approach described in \citet{Trotta22}. 

\begin{figure*}
\includegraphics[width=0.5\textwidth]{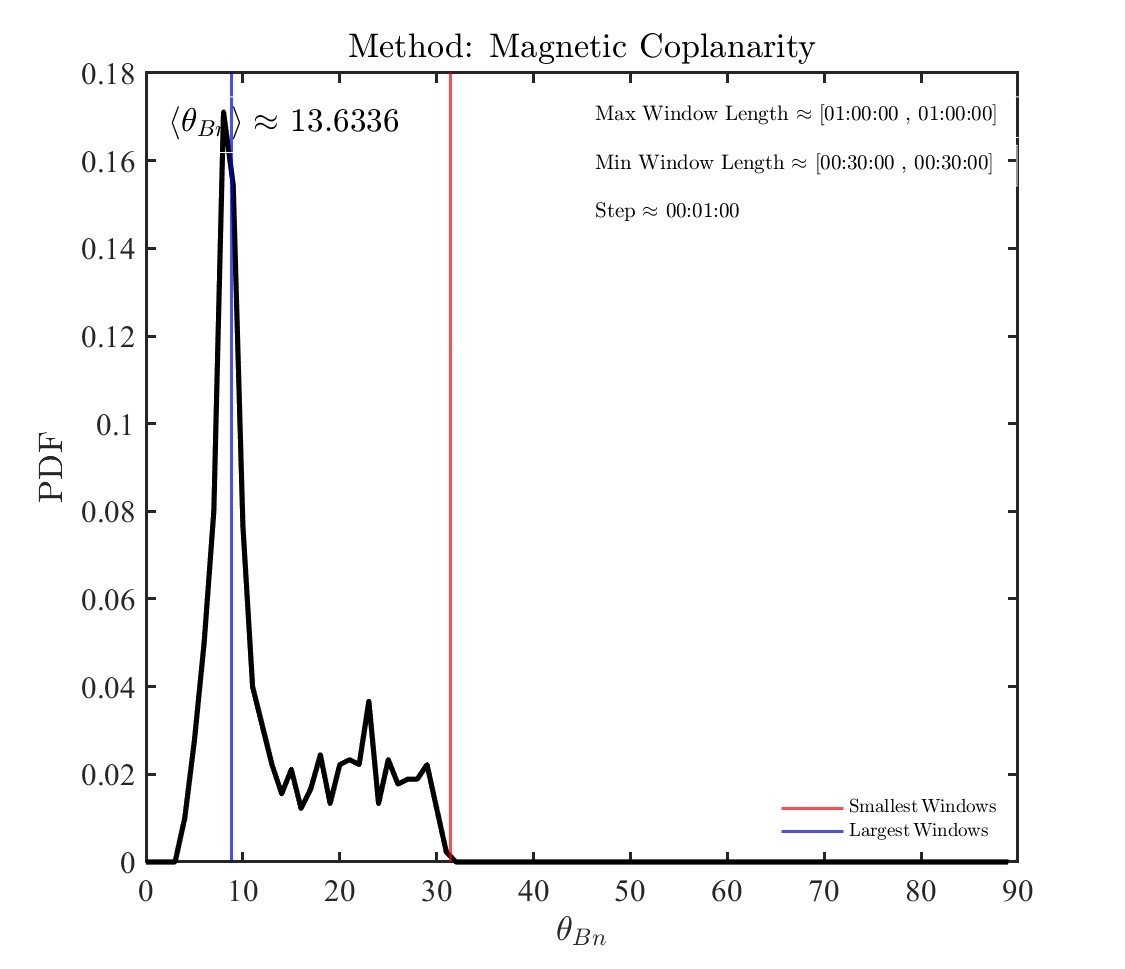}
\includegraphics[width=0.5\textwidth]{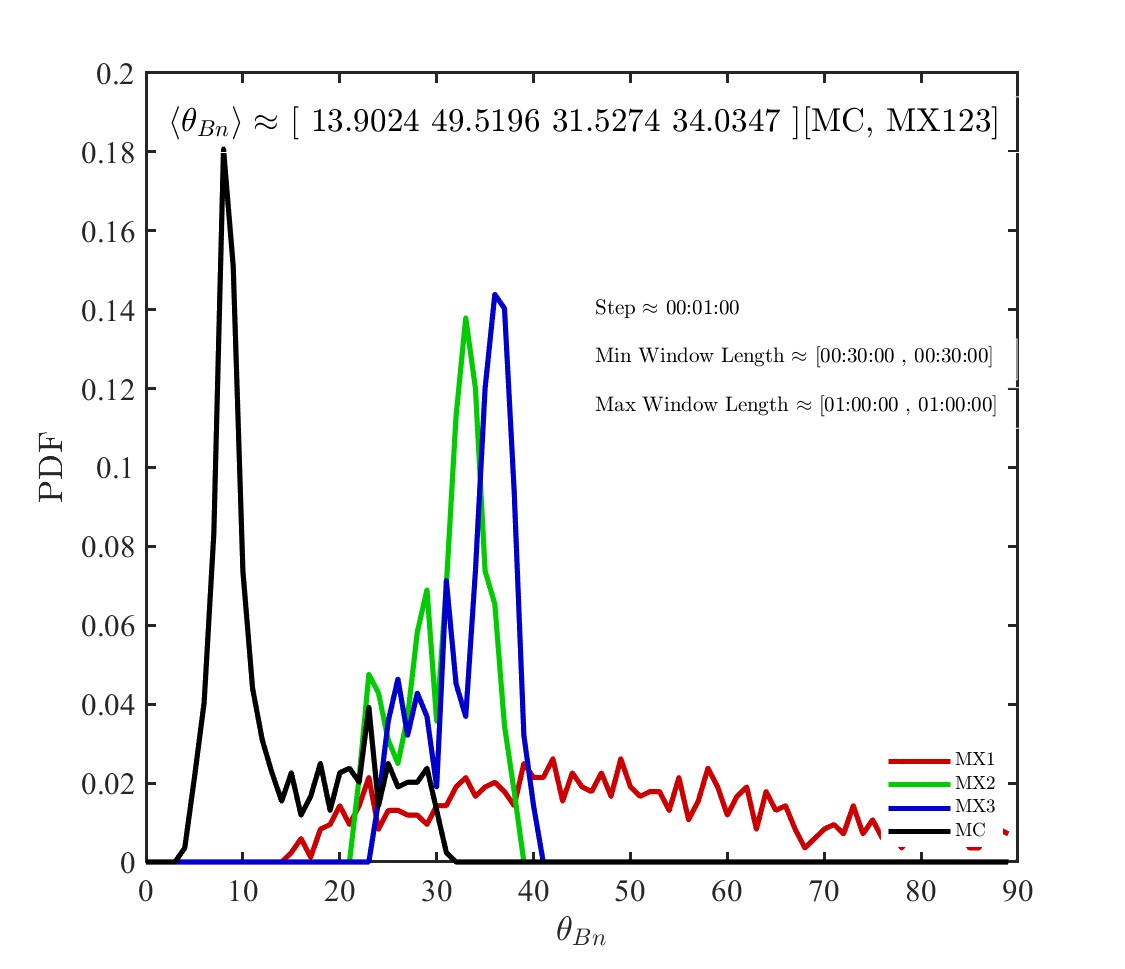}
\caption{Results of computations of $\theta_\mathrm{Bn}$ by different methods (magnetic co-planarity, mixed mode), using averaging windows varying in length (from 30~min to 1~h with $\Delta t = 1$~min). 
\label{fig:theta_bn_distributions}}
\end{figure*}

Earlier, \citet{Giacalone12} obtained for a number of ESP events an estimate of the proton acceleration time from some low (injection) energy even to $\sim$~50~keV to be of the order of an hour. This implies that small averaging windows (of the order of several minutes) should not be relevant.  
In computations of distributions of $\theta_\mathrm{Bn}$, we used averaging time windows varying in length from 30~min to 1~h with with a systematic increment $\Delta t$ of 1~min, as highlighted in Fig.~\ref{fig:shock_overview}. Figure~\ref{fig:theta_bn_distributions} shows the probability density functions (PDFs) for computed $\theta_\mathrm{Bn}$ values obtained by different methods to calculate the shock-normal vector, namely the magnetic coplanarity (MC) method and the three mixed mode (MX1, MX2 and MX3) methods \citep{Trotta22}. The MC method uses only the data on the upstream and downstream magnetic field vectors, whereas the MX methods process the bulk flow velocity data together with the magnetic field vector data \citep[for details see][]{Paschmann2000}. The left panel of Fig.~\ref{fig:theta_bn_distributions} shows the PDF obtained with the MC method. All the values of $\theta_\mathrm{Bn}$ in this case are less than $\sim 30^\circ$ with a clear peak at $\sim 8^\circ$ and $\langle\theta_\mathrm{Bn}\rangle \sim 14^\circ$. Note that smaller values of $\theta_\mathrm{Bn}$ result from larger averaging windows. The right panel of Fig.~\ref{fig:theta_bn_distributions} includes PDFs resulting from the MX1-2-3 methods. The PDFs obtained with MX2 and MX3 are quite narrow with $\langle\theta_\mathrm{Bn}\rangle \sim 32^\circ$ and $\sim 34^\circ$ correspondingly, but the PDF resulting from MX1 is much wider, with $\langle\theta_\mathrm{Bn}\rangle \sim 50^\circ$ and without a pronounced peak. One can see that all the methods provide quite different results (although three of the four methods favor the quasi-parallel shock). We note that the upstream plasma bulk speed in Fig.~\ref{fig:shock_overview} shows rather large variations seemingly associated with the current sheet crossings. This and also the significant disagreement between some versions of the MX method might be suggestive of lower reliability of the MX methods compared to the MC method in the current event. Therefore, for our simulations we still take $\theta_\mathrm{Bn} = 16^\circ$ indicated in the IP shock database of University of Helsinki, which is slightly larger than $\langle\theta_\mathrm{Bn}\rangle$ given by the MC method.          

For comparison with simulations, presented in the next section, we obtained from the particle data (Fig.~\ref{fig:esp_data_zoom_in}) the particle energy spectrum at the shock and the (omnidirectional) particle intensity vs. distance along the magnetic field in the shock upstream in different energy channels. The spectrum was derived by taking the omnidirectional intensities measured at the second minute after the shock crossing time. The particle intensities vs. distance (distance-intensity profiles) were derived from the time-intensity profiles by converting the time to the distance from the shock $x$ by applying $x = V_\mathrm{sc} (t - t_\mathrm{sh})$, where $t_\mathrm{sh}$ is the shock arrival time to the spacecraft and $V_\mathrm{sc}$ is the spacecraft speed along the upstream magnetic field relative to the shock. The latter quantity was calculated as $V_\mathrm{sc} = V_\mathrm{sh}/\cos{\theta_\mathrm{Bn}}$, where $V_\mathrm{sh} = 547$~km/s as indicated in the IP shock database of University of Helsinki.

\section{Results and discussion} \label{sec:results}
\begin{table*}[t]
    \centering
    \caption{Summary of the upstream plasma and shock parameters provided in the IP shock database of University of Helsinki and used/obtained in the SOLPACS simulations. 
    \label{table:sim_params} }
    \centering
    \begin{tabular}{l c c }
        \hline
        \hline
        Parameter & In database & In simulations\\ 
        \hline
        Ambient plasma density, $n \, [\mathrm{cm}^{-3}]$  & $3.99 \pm 1.10$  &  3.99 \\
        Upstream Alfv\'en speed, $V_\mathrm{A} \, [\mathrm{km\:s}^{-1}]$ & $66 \pm 8$ & 64 \\
        Alfv\'enic Mach number of the shock, $M^{*}_\mathrm{A}$  & $3.1 \pm 1.8$ & 3.1 \\
        Shock-normal angle, $\theta_\mathrm{Bn} \, [^\circ]$ & $16 \pm 60$ & 16 \\
        \hline
        Upstream plasma beta, $\beta$ & $0.9 \pm 0.4$ & $0.5$ \\
        Gas compression ratio of the shock, $r$ & $3.18 \pm 1.05$ & 3.34 \\
        Magnetic compression ratio of the shock, $r_\mathrm{B}$ & $1.69 \pm 0.76$ & 1.56 \\        
        \hline
    \end{tabular}
\end{table*}

The SOLPACS main input parameter values (and of the plasma beta) are given in Table~\ref{table:sim_params}. The other physical parameters specifying the initial state of the system are the following: the initial mean free path $\lambda_0$ for 100~keV protons is 0.23~au, the power-law index of the initial Alfv\'en wave spectrum $q_0$ is 1.5, the parameter $v_1$ of the seed proton spectrum given by Eq.~\eqref{eq:seed_spectrum} is taken to be 375~km/s. It should be noted that the initial mean free path in the simulations depends on the particle energy as $E^{(2-q_0)/2}$, which results from the quasi-linear theory. Specifically, for the prescribed $q_0 = 1.5$, we obtain $\lambda_0 \propto E^{1/4}$. The simulation box size is equal to 0.33~au.

The initial mean free path reference value chosen in our simulations is consistent with the values considered in SEP transport simulation studies for the ambient solar wind. For instance, in \cite{Wijsen22} the mean free path of 0.1 au for 88~keV protons was chosen, whereas our choice of values gives $\sim 0.2$~au at this energy.

It should be noted that, if the effect of self-generated waves dominates in the system (which is the case in our simulations), the influence of the exact initial mean free path is negligible. Similarly, under such conditions there is no effect of the initial wave spectral form (i.e., we could take the Kolmogorov initial spectrum with $q_0 = 5/3$). 

To constrain the last parameter of the SOLPACS input parameter list, namely the particle injection efficiency of the shock, $\epsilon_\mathrm{inj}$, we ran simulations for different values of this parameter and compared the simulated particle energy spectra at shock with the one derived from the observations. For an accurate comparison, we transformed the output simulated spectra, computed in the upstream bulk plasma frame to the spacecraft frame by applying a non-relativistic Compton-Getting correction of the omnidirectional intensity accounting for second-order terms in $V/v$ \citep[cf.][]{Forman70}: 
\begin{equation}
    I(E) \simeq I'(E) + \frac{1}{3}\left.\left(I'(E') - E' \frac{dI'}{dE'} + 2 E'^2 \frac{d^2I'}{dE'^2} \right)\right|_{E' = E} \frac{mV^{2}}{2E},
    \label{eq:compton_getting_correction}
\end{equation}
where $I(E) = p^2 f^{(\mathrm{sc})}_0(p)$ is the omnidirectional particle intensity in the spacecraft frame as a function of energy $E = mv^2/2$, $p$ is the particle momentum magnitude, $f^{(\mathrm{sc})}_0(p)$ is the isotropic part of the distribution function in this frame, $I'(E')$ is the omnidirectional intensity in a different (e.g., upstream plasma) frame,  and $V$ is the relative speed of one frame wrt. the other one. The derivation is presented in the Appendix. It is easy to see that the omnidirectional intensity is invariant to first order in $V/v$ \citep{Forman70}.

We applied the above correction, using an estimation for the relative frame speed (between the upstream plasma frame and the spacecraft frame), $V \approx V_\mathrm{sh} - u_1 = V_\mathrm{sh} - M_\mathrm{A} V_\mathrm{A} = 334$~km/s, neglecting the small nonalignment of the ambient magnetic field, ambient solar wind and shock normal vectors. The correction is small at the observed energies and can be neglected at $E > 200$~keV.

\begin{figure*}
\centering
\includegraphics[width=0.48\textwidth]{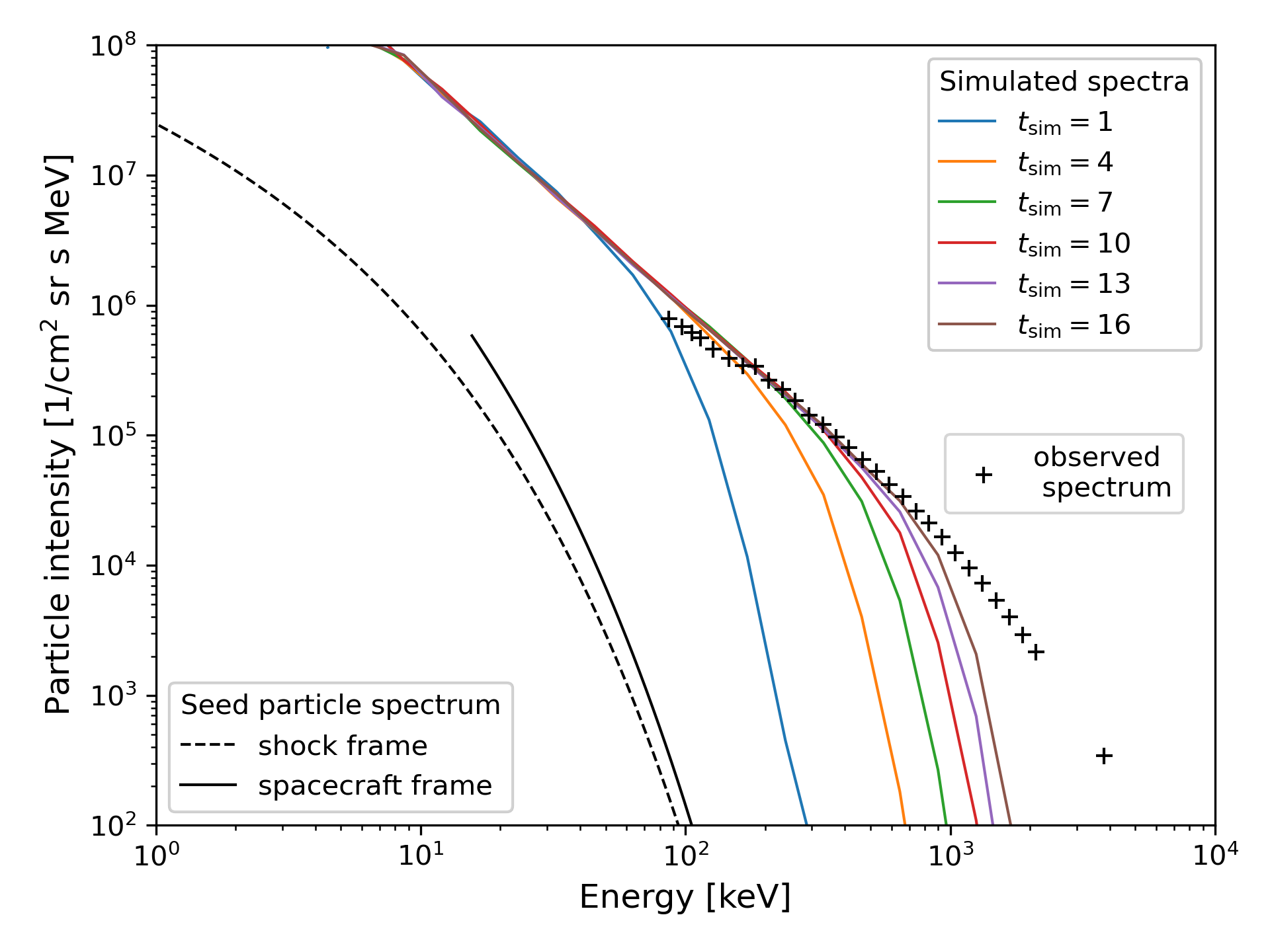}
\includegraphics[width=0.48\textwidth]{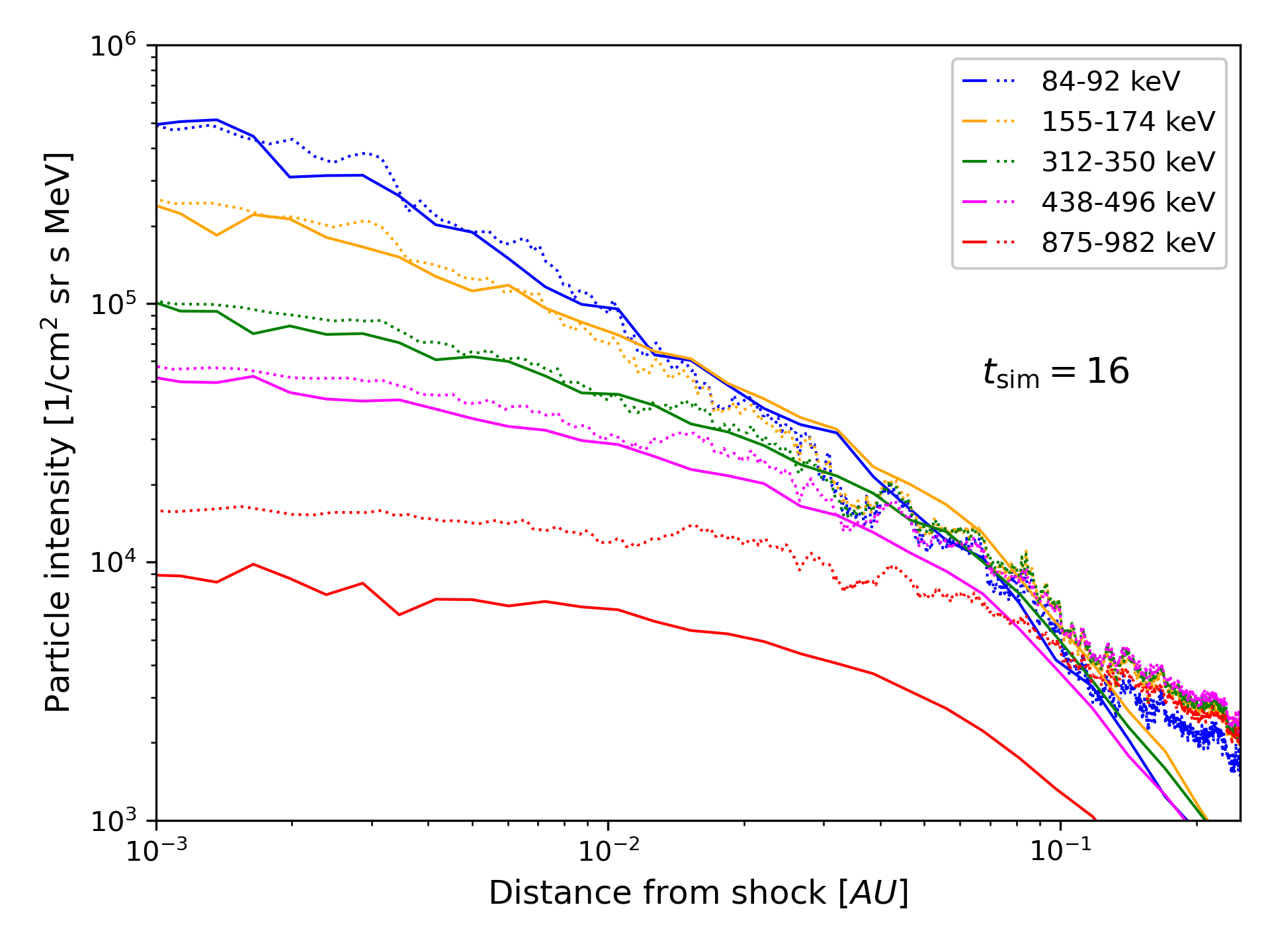}
\caption{Comparison of simulations with observations.
Left panel: Time evolution of the proton energy spectrum at the shock in the simulation for $\epsilon_\mathrm{inj} = 6 \cdot 10^{-3}$ (colored lines) superposed on the spectrum obtained from the observations (crosses). The indicated simulation times are in arbitrary units. Also shown is the seed particle spectrum as computed in the shock frame (black dashed line) and in the spacecraft frame (solid black line). See text for further details.
Right panel: Simulated proton distributions along the magnetic field line at $t_\mathrm{sim} = 16$ [arbitrary units] (solid lines) vs. the distributions obtained from the observations (dotted lines). 
\label{fig:comparison}}
\end{figure*}

Figure~\ref{fig:comparison}, left panel shows the time evolution of the spectrum in the simulation for $\epsilon_\mathrm{inj} = 6 \cdot 10^{-3}$ superposed on the observed spectrum. Also shown is the injected seed particle spectrum in the shock frame and in the spacecraft frame. The latter spectrum is computed using Eq.~\eqref{eq:compton_getting_correction} with $V = V_\mathrm{sh} = 547$~km/s. Note that all the spacecraft-frame spectra are shown only at those energies where $V/v \lesssim 0.3$. One can see that the evolving spectrum arrives to a steady state at energies below the roll-over energy, which matches well with the observed spectrum. We should note that the obtained value of $\epsilon_\mathrm{inj}$ is close to a few percent of solar wind protons reflected by quasi-parallel shocks as reported in previous studies \citep[see, e.g.,][]{NgReames08}.

Having fixed $\epsilon_\mathrm{inj}$ as described above, we computed the particle distance-intensity profiles in this simulation and compared those with the profiles derived from the observations (Fig.~\ref{fig:comparison},  right panel). One can see an excellent match at those energies where the spectra match as well. This result, i.e., the correspondence obtained simultaneously for both the energy spectrum and the distance-intensity profiles of protons for the same set of values of the input parameters, strongly suggests that it is the process of self-generation of Alfv\'en waves that controls the proton energization and mean free path in the upstream near the shock in this ESP event, rather than, for example, super-diffusive transport of particles. 

At the same time, it can be seen that there is no match between the simulations and observations at the highest energy channel 875--982 keV, in which one can still see an enhancement in intensity (Fig.~\ref{fig:esp_data_zoom_in}). However, the particle population at these (and higher) energies is not in a steady state by the end of the simulation, but it will tend to increase, if the simulation is continued. 
Note in this connection that the simulation times, $t_\mathrm{sim}$, given in Fig.~\ref{fig:comparison} in arbitrary units could be given in physical units, e.g., hours. However, since the initial state of the modeled system in the simulations is characterized by a very low turbulence level, the acceleration time to a certain energy in the simulations is incomparably longer than the acceleration time estimated from the observed time-intensity profiles in ESP events, using classical DSA theory \citep[for estimates of the acceleration times from ESP event observations see, e.g.,][]{Giacalone12}. This is because such estimates are done under the assumption of steady particle diffusion coefficients, while in the simulation it takes additional time for the accelerated particles to amplify the waves at resonant frequencies and bring them to a steady state. 
Taking that into consideration, the time in our simulations in the context of ESP event modeling is a parameter rather than the physical time. This makes it difficult to conclude whether the insufficient simulation time is the reason for the mismatch. 

Another possible reason for that may be the spatial locality of the simulation (constant plasma and shock parameters). It may take $\sim 10$~h to accelerate protons up to 1~MeV \citep[extrapolating the acceleration times plotted in Fig.~10 in][]{Giacalone12}, so the constancy of the parameters can be quite a limiting approximation for high-energy particles.     
Note that the time-intensity profiles at energies $>1.8$~MeV (Fig.~\ref{fig:esp_data_zoom_in}) do not peak at the shock, but show a quasi-plateau-like shape with a smeared maximum before the shock arrival. This indicates that the shock does not accelerate particles at these energies at the time of passage of the spacecraft, but it did so earlier, so the shock was stronger at shorter heliocentric distances. 
 
\section{Conclusions and Outlook} \label{sec:conclusions}
In this study, we have addressed the question on how well the modeling of particle (proton) acceleration in shocks, based on self-consistent quasi-linear description of wave-particle interactions, can reproduce actual observations. To answer this question, we simulated the ESP event of November 10, 2012 observed by STEREO-A with the SOLPACS code \citep{Afanasiev15} upgraded to be applicable to oblique shocks. We compared the proton energy spectrum at the shock and proton intensity distributions in the upstream obtained in the simulations with those derived from the observations. We found excellent correspondence for the simulation input parameters fixed by the in-situ plasma measurements at the spacecraft, at the energies at which the simulated system comes to a steady state by the end of the simulation. The mismatch at higher energies (at which the system is not in a steady state by the end of a simulation) can be due to too short simulation time or due to spatial locality of SOLPACS.  
Nevertheless, our study strongly supports the idea that the wave growth due to streaming particles is involved in the particle acceleration process in shocks. It also supports the quasi-linear description of wave-particle interactions, being the core of many self-consistent models of particle acceleration in shocks, and validates the SOLPACS code. It is of interest to model ESP events measured closer to the Sun. Currently, we are simulating several ESP events observed by the Solar Orbiter spacecraft at heliocentric distances $< 1$~au. Preliminary, we are having a good correspondence between the simulations and the observations as well. This will be addressed in detail in a separate study. 
Another possibility that will be explored is modeling of the particle injection spectrum as consisting of not only the supra-thermal component, but also an energetic component in order to mimic the preceding SEP event. This may lead to a better match with observations at energies above the spectral roll-over.


\begin{acknowledgements}
This research has received funding from the European Union’s Horizon 2020 research and innovation programme under grant agreements No 870405 (EUHFORIA 2.0) and 101004159 (SERPENTINE). The work in the University of Turku is performed under the umbrella of Finnish Centre of Excellence in Research of Sustainable Space (FORESAIL) funded by the Academy of Finland (grant No. 336809). N.D. is grateful for support by the Academy of Finland (SHOCKSEE, grant No.\ 346902). We also acknowledge the computer resources of the Finnish IT Center for Science (CSC) and the FGCI project (Finland).
\end{acknowledgements}

\bibliographystyle{aa}
\bibliography{references}

\begin{appendix}
\onecolumn
\section{Second-order Compton-Getting correction of the particle omnidirectional intensity}
Let us denote in this section the particle six-dimensional phase-space distribution function in the observer's (primed) frame to be $f'$ and assume the distribution $f$ in the unprimed frame to be isotropic. We also assume that the unprimed frame moves wrt. the primed one (the observer) with velocity $\mathbf{V}$. Then due to the Lorentz-invariance of the distribution function \citep{Forman70}, we can write:

\begin{eqnarray*}
f'(\mathbf{p}') & = & f(p)\\
 & = & f\left(\sqrt{p_{x}^{2}+p_{y}^{2}+p_{z}^{2}}\right)\\
 & = & f(p')+\left.\left(\frac{df}{dp}\frac{\partial p}{\partial p_{x}}\right)\right|_{\mathbf{p}=\mathbf{p}'}\left(p_{x}-p'_{x}\right)+\left.\left(\frac{df}{dp}\frac{\partial p}{\partial p_{y}}\right)\right|_{\mathbf{p}=\mathbf{p}'}\left(p_{y}-p'_{y}\right)+\left.\left(\frac{df}{dp}\frac{\partial p}{\partial p_{z}}\right)\right|_{\mathbf{p}=\mathbf{p}'}\left(p_{z}-p'_{z}\right)\\
 &  & +\frac{1}{2}\left.\frac{\partial^{2}f}{\partial p_{x}^{2}}\right|_{\mathbf{p}=\mathbf{p}'}\left(p_{x}-p'_{x}\right)^{2}+\frac{1}{2}\left.\frac{\partial^{2}f}{\partial p_{y}^{2}}\right|_{\mathbf{p}=\mathbf{p}'}\left(p_{y}-p'_{y}\right)^{2}+\frac{1}{2}\left.\frac{\partial^{2}f}{\partial p_{z}^{2}}\right|_{\mathbf{p}=\mathbf{p}'}\left(p_{z}-p'_{z}\right)^{2}\\
 &  & +\left.\frac{\partial^{2}f}{\partial p_{x}\partial p_{y}}\right|_{\mathbf{p}=\mathbf{p}'}\left(p_{x}-p'_{x}\right)\left(p_{y}-p'_{y}\right)+\left.\frac{\partial^{2}f}{\partial p_{x}\partial p_{z}}\right|_{\mathbf{p}=\mathbf{p}'}\left(p_{x}-p'_{x}\right)\left(p_{z}-p'_{z}\right)\\
 &  & +\left.\frac{\partial^{2}f}{\partial p_{z}\partial p_{y}}\right|_{\mathbf{p}=\mathbf{p}'}\left(p_{z}-p'_{z}\right)\left(p_{y}-p'_{y}\right)+O\left(\left|\mathbf{p}-\mathbf{p}'\right|^{3}\right),
\end{eqnarray*}
where $\mathbf{p}$ and $\mathbf{p}'$ denote the particle momentum in the respective frame, and we make use of Cartesian coordinates in the momentum space. Neglecting for the moment the second- and higher order terms, we obtain
\begin{eqnarray*}
f'(\mathbf{p}') & \simeq & f(p')+\left.\frac{df}{dp}\right|_{p=p'}\frac{p'_{x}}{p'}\left(p_{x}-p'_{x}\right)+\left.\frac{df}{dp}\right|_{p=p'}\frac{p'_{y}}{p'}\left(p_{y}-p'_{y}\right)+\left.\frac{df}{dp}\right|_{p=p'}\frac{p'_{z}}{p'}\left(p_{z}-p'_{z}\right)\\
 & = & f(p')+\left.\frac{df}{dp}\right|_{p=p'}\frac{\mathbf{p}'}{p'}\boldsymbol{\cdot}(\mathbf{p}-\mathbf{p}')\\
 & = & f(p')-\left.\frac{df}{dp}\right|_{p=p'}\mathbf{n}'\boldsymbol{\cdot}\frac{p'}{v'}\mathbf{V},
\end{eqnarray*}
where we used $\mathbf{p}-\mathbf{p}'=-m\mathbf{V}$ (i.e., non-relativistic consideration), and $\mathbf{n}'=\mathbf{p}'/p'$ is a unit vector in the direction of the particle's momentum. Thus, we reproduce \citeauthor{Forman70}'s \citeyearpar{Forman70} result.

The second-order terms are
\begin{align*}
\frac{\partial^{2}f}{\partial p_{x}^{2}} & =\frac{\partial}{\partial p_{x}}\left(\frac{\partial f}{\partial p_{x}}\right)=\frac{\partial}{\partial p_{x}}\left(\frac{df}{dp}\frac{\partial p}{\partial p_{x}}\right)=\frac{\partial}{\partial p_{x}}\left(\frac{df}{dp}\frac{p_{x}}{p}\right)
= \frac{d^{2}f}{dp^{2}}\left(\frac{p_{x}}{p}\right)^{2}+\frac{df}{dp}\left(\frac{1}{p}-\frac{p_{x}^{2}}{p^{3}}\right)=\frac{d^{2}f}{dp^{2}}\left(\frac{p_{x}}{p}\right)^{2}+\frac{1}{p}\frac{df}{dp}\left(1-\frac{p_{x}^{2}}{p^{2}}\right),
\end{align*}
\[
\frac{\partial^{2}f}{\partial p_{y}^{2}}=\frac{d^{2}f}{dp^{2}}\left(\frac{p_{y}}{p}\right)^{2}+\frac{1}{p}\frac{df}{dp}\left(1-\frac{p_{y}^{2}}{p^{2}}\right),
\]
\[
\frac{\partial^{2}f}{\partial p_{z}^{2}}=\frac{d^{2}f}{dp^{2}}\left(\frac{p_{z}}{p}\right)^{2}+\frac{1}{p}\frac{df}{dp}\left(1-\frac{p_{z}^{2}}{p^{2}}\right),
\]
\[
\frac{\partial^{2}f}{\partial p_{x}\partial p_{y}}=\frac{\partial}{\partial p_{x}}\left(\frac{\partial f}{\partial p_{y}}\right)=\frac{\partial}{\partial p_{x}}\left(\frac{df}{dp}\frac{p_{y}}{p}\right)=\frac{d^{2}f}{dp^{2}}\frac{p_{x}p_{y}}{p^{2}}-\frac{df}{dp}\frac{p_{y}p_{x}}{p^{3}}=\frac{p_{x}p_{y}}{p^{2}}\left(\frac{d^{2}f}{dp^{2}}-\frac{1}{p}\frac{df}{dp}\right),
\]
\[
\frac{\partial^{2}f}{\partial p_{x}\partial p_{z}}=\frac{p_{x}p_{z}}{p^{2}}\left(\frac{d^{2}f}{dp^{2}}-\frac{1}{p}\frac{df}{dp}\right),
\]
\[
\frac{\partial^{2}f}{\partial p_{z}\partial p_{y}}=\frac{p_{z}p_{y}}{p^{2}}\left(\frac{d^{2}f}{dp^{2}}-\frac{1}{p}\frac{df}{dp}\right).
\]

Introducing the spherical coordinates
\begin{align*}
p_{x} & =  p\sin\theta\cos\phi,\\
p_{y} & =  p\sin\theta\sin\phi,\\
p_{z} & =  p\cos\theta,
\end{align*}
the second-order terms can be given as
\[
\frac{\partial^{2}f}{\partial p_{x}^{2}}=\frac{d^{2}f}{dp^{2}}\sin^{2}\theta\cos^{2}\phi+\frac{1}{p}\frac{df}{dp}\left(1-\sin^{2}\theta\cos^{2}\phi\right),
\]
\[
\frac{\partial^{2}f}{\partial p_{y}^{2}}=\frac{d^{2}f}{dp^{2}}\sin^{2}\theta\sin^{2}\phi+\frac{1}{p}\frac{df}{dp}\left(1-\sin^{2}\theta\sin^{2}\phi\right),
\]
\[
\frac{\partial^{2}f}{\partial p_{z}^{2}}=\frac{d^{2}f}{dp^{2}}\cos^{2}\theta+\frac{1}{p}\frac{df}{dp}\left(1-\cos^{2}\theta\right),
\]
\[
\frac{\partial^{2}f}{\partial p_{x}\partial p_{y}}=\left(\frac{d^{2}f}{dp^{2}}-\frac{1}{p}\frac{df}{dp}\right)\sin^{2}\theta\sin\phi\cos\phi,
\]
\[
\frac{\partial^{2}f}{\partial p_{x}\partial p_{z}}=\left(\frac{d^{2}f}{dp^{2}}-\frac{1}{p}\frac{df}{dp}\right)\cos\theta\sin\theta\cos\phi,
\]
\[
\frac{\partial^{2}f}{\partial p_{z}\partial p_{y}}=\left(\frac{d^{2}f}{dp^{2}}-\frac{1}{p}\frac{df}{dp}\right)\cos\theta\sin\theta\sin\phi.
\]

Averaging the second-order terms over angles, we get
\[
\left\langle \left.\frac{\partial^{2}f}{\partial p_{x}^{2}}\right|_{\mathbf{p}=\mathbf{p}'}\left(p_{x}-p'_{x}\right)^{2}\right\rangle _{\theta,\phi}=\left.\left(\frac{1}{3}\frac{d^{2}f}{dp^{2}}+\frac{2}{3p}\frac{df}{dp}\right)\right|_{p=p'}m^{2}V_{x}^{2},
\]
\[
\left\langle \left.\frac{\partial^{2}f}{\partial p_{y}^{2}}\right|_{\mathbf{p}=\mathbf{p}'}\left(p_{y}-p'_{y}\right)^{2}\right\rangle _{\theta,\phi}=\left.\left(\frac{1}{3}\frac{d^{2}f}{dp^{2}}+\frac{2}{3p}\frac{df}{dp}\right)\right|_{p=p'}m^{2}V_{y}^{2},
\]
\[
\left\langle \left.\frac{\partial^{2}f}{\partial p_{z}^{2}}\right|_{\mathbf{p}=\mathbf{p}'}\left(p_{z}-p'_{z}\right)^{2}\right\rangle _{\theta,\phi}=\left.\left(\frac{1}{3}\frac{d^{2}f}{dp^{2}}+\frac{2}{3p}\frac{df}{dp}\right)\right|_{p=p'}m^{2}V_{z}^{2},
\]
\begin{align*}
\left\langle \left.\frac{\partial^{2}f}{\partial p_{x}\partial p_{y}}\right|_{\mathbf{p}=\mathbf{p}'}\left(p_{x}-p'_{x}\right)\left(p_{y}-p'_{y}\right)\right\rangle _{\theta,\phi} =\left\langle \left.\frac{\partial^{2}f}{\partial p_{x}\partial p_{z}}\right|_{\mathbf{p}=\mathbf{p}'}\left(p_{x}-p'_{x}\right)\left(p_{z}-p'_{z}\right)\right\rangle _{\theta,\phi}
= \left\langle \left.\frac{\partial^{2}f}{\partial p_{z}\partial p_{y}}\right|_{\mathbf{p}=\mathbf{p}'}\left(p_{z}-p'_{z}\right)\left(p_{y}-p'_{y}\right)\right\rangle _{\theta,\phi}=0.
\end{align*}

Taking $V_{x}=V_{y}=0$ and $V_{z}=V$, we obtain for the isotropic
part of the distribution function in the observer's frame:
\begin{eqnarray*}
\left\langle f'(\mathbf{p}')\right\rangle _{\theta,\phi} & = & f(p')+\frac{1}{2}\left.\left(\frac{1}{3}\frac{d^{2}f}{dp^{2}}+\frac{2}{3p}\frac{df}{dp}\right)\right|_{p=p'}m^{2}V^{2}+o\left(\frac{m^{2}V^{2}}{p'^{2}}\right)\\
 & = & f(p')+\left.\left(\frac{1}{6}p{}^{2}\frac{d^{2}f}{dp^{2}}+\frac{1}{3}p\frac{df}{dp}\right)\right|_{p=p'}\frac{m^{2}V^{2}}{p'^{2}}+o\left(\frac{m^{2}V^{2}}{p'^{2}}\right).
\end{eqnarray*}

Neglecting the terms higher than $m^2V^2/p'^2$, we obtain for the omnidirectional differential intensity in the observer's frame,
$I'(E')$: 
\begin{eqnarray*}
I'(E') & = & p'^{2}\left\langle f'(\mathbf{p}')\right\rangle _{\theta,\phi}\\
 & \simeq & p'^{2}f(p')+p'^{2}\left.\left(\frac{1}{6}p{}^{2}\frac{d^{2}f}{dp^{2}}+\frac{1}{3}p\frac{df}{dp}\right)\right|_{p=p'}\frac{m^{2}V^{2}}{p'^{2}}\\
 & = & I(E') + \frac{1}{3}\left.\left(I(E) - E \frac{dI}{dE} + 2 E^2 \frac{d^2I}{dE^2} \right)\right|_{E = E'} \frac{mV^{2}}{2E'},
\end{eqnarray*}
where $I(E)$ is the omnidirectional intensity in the unprimed frame, being a function of energy $E$.

To ease the notation in the main body of this paper, let us swap the primed and unprimed quantities in the resulting expression to get:
\[
I(E) \simeq I'(E) + \frac{1}{3}\left.\left(I'(E') - E' \frac{dI'}{dE'} + 2 E'^2 \frac{d^2I'}{dE'^2} \right)\right|_{E' = E} \frac{mV^{2}}{2E},
\]
which coincides with Eq.~\eqref{eq:compton_getting_correction}.
\end{appendix}

\end{document}